\documentclass[twocolumn, prl, showpacs]{revtex4-1}

\usepackage{graphicx}
\usepackage{dcolumn}
\usepackage{bm}
\usepackage{physics}
\usepackage{verbatim}
\usepackage{amsmath}
\usepackage{mathtools}
\usepackage{placeins}
\usepackage[usenames,dvipsnames]{color}

\definecolor{aaltoOrange}{RGB}{255,121,0}%

\begin{document}

\title{A quantum embedding theory in the screened Coulomb interaction: Combining configuration interaction with \textit{GW}\hspace{-0.1cm}/BSE}

\author{Marc Dvorak}
\email{marc.dvorak@aalto.fi}
\affiliation{Department of Applied Physics, Aalto University School of Science, 00076-Aalto, Finland}
\author{Dorothea Golze}
\affiliation{Department of Applied Physics, Aalto University School of Science, 00076-Aalto, Finland}
\author{Patrick Rinke}
\affiliation{Department of Applied Physics, Aalto University School of Science, 00076-Aalto, Finland}

\date{\today}

\newcommand{\sd}{$\mathcal{D}\;$}
\newcommand{\sr}{$\mathcal{R}\;$}
\newcommand{\md}{\mathcal{D}}
\newcommand{\mr}{\mathcal{R}}

\newcommand{\id}{\mathbf{I}}

\begin{abstract}
We present a new quantum embedding theory called dynamical configuration interaction (DCI) that combines wave function and Green's function theories. DCI captures static correlation in a correlated subspace with configuration interaction and couples to high-energy, dynamic correlation outside the subspace with many-body perturbation theory based on Green's functions. DCI takes the strengths of both theories to balance static and dynamic correlation in a single, fully \textit{ab-initio} embedding concept. The theory adds dynamic correlation around a fixed active space of orbitals with efficient $\mathcal{O}(N^5)$ scaling, while maintaining a multi-reference treatment of the active space. We show that treating high-energy correlation up to the $GW$ and Bethe-Salpeter equation level is sufficient even for challenging multi-reference problems. Our theory treats ground and excited states on equal footing, and we compute the dissociation curve of N$_2$, vertical excitation energies of small molecules, and the ionization spectrum of benzene in excellent agreement with high level quantum chemistry methods and experiment.
\end{abstract}

\maketitle

The quantum many-body problem is at the heart of chemical reactions, emergent phenomena in materials, and countless technological applications. Consequently, the prediction of ground and excited states of quantum many-body systems remains one of the most intensely researched topics in physics, materials science, and chemistry. The diversity of the quantum many-body problem arises from the dramatic variation of electronic correlation: from the highly multi-reference character along reaction pathways in quantum chemistry to dynamical screening in polarizable materials. Theories from different disciplines describe certain regimes of correlation better than others, with widely varying computational costs \cite{zgid_prx_7}. Accordingly, there is great potential for new methods which combine theories to enhance their respective strengths and downplay their weaknesses.

In this article, we highlight a new quantum embedding theory to merge complementary disciplines. In an active space (AS) of strongly-correlated orbitals, we diagonalize the many-body Hamiltonian with the configuration interaction (CI) approach. In addition to the interaction between these strongly-correlated configurations, we downfold the effects of high-energy transitions onto an energy-dependent correction added to the CI Hamiltonian. We estimate these dynamical corrections with a modified $GW$ plus Bethe-Salpeter equation (BSE) procedure. Our energy-dependent corrections correlate the full set of orbitals \textit{beyond} the orbital AS and add dynamic correlation from the bath with only $\mathcal{O}(N^5)$ scaling.

Quantum embedding or AS theories which reduce the effective size of the Hamiltonian are not a new idea in strongly-correlated physics and quantum chemistry \cite{chan_acr_49}. However, fully \textit{ab-initio} embedding theories that are still computationally feasible are difficult to formulate. Exact embedding frameworks exist \cite{aryasetiawan_prl_102,lowdin_jmp_3} but, without any simplification, are essentially as intractable as the initial many-body problem. Approximate model Hamiltonians \cite{hubbard,bockstedte_npjqm_3,ivady_prb_90} are useful to reduce the computational cost but may rely on semi-empirical or otherwise not \textit{ab-initio} parameters. Computationally feasible, \textit{ab-initio} embedding theories have proven to be extremely valuable for studying strongly-correlated systems \cite{biermann_jpcm_26,tomczak_epl_100,zgid_prb_91,rusakov_jctc_15,chan_prl_109,knizia_jctc_9,casula_prl_109,olsen_ijqc_111,pulay_ijqc_111,loos_jcp_149,knizia_prl_109}.

Different many-body methods have distinct advantages. Exact diagonalization (ED) of the many-body Hamiltonian describes all static correlation or multi-reference character in a frequency independent framework \cite{helgaker_molecular}. ED suffers from a combinatorial explosion in the basis, but its truncated basis version configuration interaction (CI) reduces to polynomial scaling. Other wave function methods, such as coupled cluster (CC), have polynomial scaling ($\mathcal{O}(N^6)$ for single and double excitations, $\mathcal{O}(N^8)$ with triple excitations), but can not necessarily treat all types of strong correlation that appear in bond breaking or open-shell problems. An alternative approach to the electronic problem is many-body perturbation theory (MBPT) \cite{martin_reining_ceperley_2016,fetter_quantum} based on Green's functions. In particular, the $GW$ approximation \cite{hedin_pr_139,gw_review} and its extension to the Bethe-Salpeter equation \cite{salpeter_pr_84} (BSE) are very successful at predicting quasiparticle excitations in weakly- to moderately-correlated materials \cite{louie_prb_34,rohlfing_prb_62,rinke_jctc_11,jacquemin_jctc_11,bruneval_jcp_142,chelikowsky_prb_73,botti_jctc_10}, with $GW$ scaling as $\mathcal{O}(N^4)$. Our motivation is to treat static correlation in a strongly-correlated subspace with CI and the remaining high-energy degrees of freedom with $GW$/BSE.

\begin{figure*}
\begin{centering}
\includegraphics[width=\textwidth]{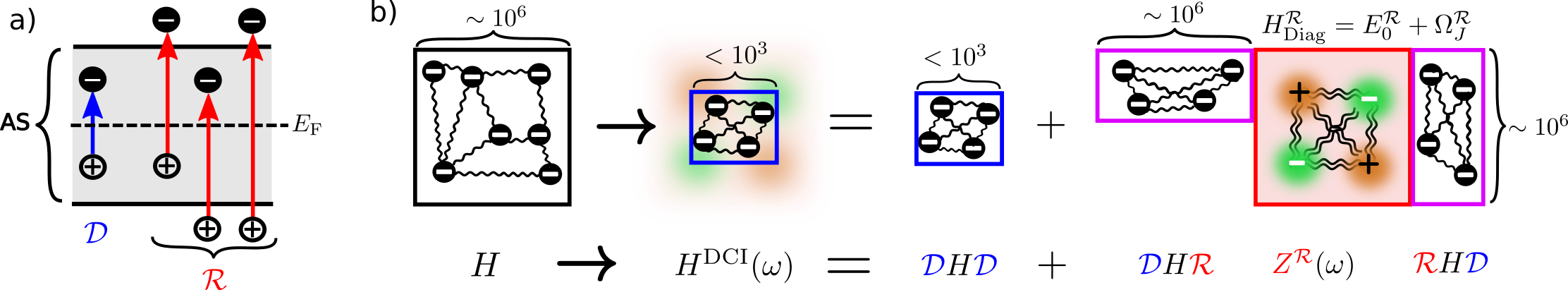}
\caption{ a) Partitioning of the many-body Hilbert space into \sd (blue) and $\mr$ (red). All excitations that fall inside the orbital AS, shaded in grey, belong to $\md$. All other configurations are placed in $\mr$. b) Matrix elements of the exact $H$ describe $N$ interacting bare electrons (black) in the vacuum (white background). The renormalized $H^{\mr}$ describes $2m$ ($m=\;$excitation level) interacting quasiparticles (orange and green) above a correlated ground state (red background). \label{embedding}}
\end{centering}
\end{figure*}

Here, we only sketch the theory and refer to Ref.~\onlinecite{dvorak_prb} for details. We consider the electronic Hamiltonian in the Born-Oppenheimer approximation,
\begin{equation}
H = \sum_{ij}^N t_{ij} a_i^{\dagger} a_j + \sum_{ijkl}^N v_{ijkl} a^{\dagger}_i a^{\dagger}_j a_l a_k . \label{secondquant}
\end{equation}
$t_{ij}$ and $v_{ijkl}$ are the one- and two-body (Coulomb) matrix elements of the Hamiltonian and $a_i$ ($a^{\dagger}_j$) are fermionic destruction (creation) operators. We divide the $N$-particle many-body Hilbert space into two portions defined by the projection operators \sd and $\mr$,
\begin{equation}
\md = \sum_{I} \ket{I} \bra{I}  \;\; ; \;\; \mr = \sum_{J} \ket{J} \bra{J} \;\; ; \;\;   \id = \md + \mr.
\end{equation}
Here, $\ket{I}$ and $\ket{J}$ are many-body configurations. To connect the many-body projectors to the single-particle picture, we define an orbital AS around the Fermi energy, as shown in Fig.~\ref{embedding}a. The AS contains the statically correlated single-particle states. We place all many-body configurations $\ket{I}$ containing AS excitations in the strongly-correlated space, $\md$. This criterion includes all excitation levels (single, double, etc.). We place all other configurations $\ket{J}$ in the weakly-correlated space, $\mr$. Based on the projectors, the Schr\"odinger equation can be downfolded onto a non-linear effective Hamiltonian in $\md$ \cite{lowdin_jmp_3,lowdin_ijqc_2,dzuba_pra_54,pavlyukh_prb_91,dzuba_pra_95,gagliardi_jcp_134,gagliardi_jctc_9,marini_jcp_134,romaniello_jcp_130}:
\begin{eqnarray}
Z^{\mr}(E) &=& \frac{1}{E - \mr H \mr}  \label{effective}  \nonumber  \\
M(E) &\equiv& \left[ \md H \mr \right] Z^{\mr}(E) \left[ \mr H \md \right]  \nonumber  \\
H^{\textrm{eff}}(E) \; \phi &=& \left[ \md H \md + M(E) \right] \phi = E \phi  \label{eigen}.   
\end{eqnarray}
 Eq.~\ref{eigen} requires inversion of the enormous matrix $\mr H \mr$, which is easily $\gg 10^{10}$ for realistic systems.

Our theory transforms the projected $\mr H \mr$ Hamiltonian to simplify the matrix inversion. By introducing a ground state energy in the $\mr$ subspace, which we denote $E_0^{\mr}$, we renormalize the subspace Hamiltonian $\mr H \mr$ to a Hamiltonian of excitations propagating over a correlated ground state. We rewrite $\mr H \mr$ as 
\begin{equation}
\mr H \mr \rightarrow H^{\mr} \equiv E_0^{\mr} + \Omega^{\mr} \label{transformation}
\end{equation}
for some ground state energy $E_0^{\mr}$ and excitation matrix $\Omega^{\mr}$. $E_0^{\mr}$ and $\Omega^{\mr}$ require a careful construction that is detailed in Ref.~\onlinecite{dvorak_prb}.

The most important aspect of our theory is the calculation of excitation energies ($\Omega^{\mr}$). To calculate $\Omega^{\mr}$, we switch from the wave function to quasiparticle picture, as dictated by the transformation in Eq.~\ref{transformation}. This transformation allows us to take advantage of the highly successful $GW$ approximation, which excels at treating dynamically correlated electrons. To lower the expense of inverting $\mr H \mr$, we adopt a diagonal approximation to $\Omega^{\mr}$. The ensuing inversion of the diagonal matrix is trivial and still correlates the full set of orbitals at the quasiparticle level.

In our quasiparticle estimate of excitation energies, the diagonal matrix elements of $\Omega^{\mr}$ are
\begin{eqnarray}
\Omega_J^{\mr} &=& \bra{J} \Omega^{\mr} \ket{J}  \nonumber  \\
&=& \sum_{e \in J}^m \epsilon_e^{GW_{\mr}} - \sum_{h \in J}^m \epsilon_h^{GW_{\mr}}   \nonumber  \\ 
&+& \sum_{e,h \in J}^m (-W_{\mr,eheh} + \delta_{\sigma_e \sigma_h} v_{ehhe} ) \nonumber  \\
&+& \sum_{\mathclap{\substack{e \in J \\
e \neq e'}}}^m \;\; (W_{\mr,ee'ee'} - \delta_{\sigma_e \sigma_{e'}} W_{\mr, ee'e'e} ) \nonumber \\  
&+& \sum_{\mathclap{\substack{h \in J \\
h \neq h'}}}^m \;\; (W_{\mr,hh'hh'} - \delta_{\sigma_h \sigma_{h'}} W_{\mr, hh'h'h} ).   \label{excitation}
\end{eqnarray}
In Eq.~\ref{excitation}, $e$ and $h$ denote electrons and holes in configuration $\ket{J}$, $\sigma$ is a spin variable, and sums run up to the excitation level $m$ of the configuration. A critical element of the construction is our use of the constrained random phase approximation (cRPA). Instead of calculating the polarizability with all single excitations, which is the normal case, the cRPA omits low energy single excitations which belong to $\md$. Screening of the bare Coulomb interaction by this constrained polarizability gives the partially screened Coulomb interaction, $W_{\mr}$. $W_{\mr}$ includes only high energy screening channels $-$ intra-$\mr$ correlation $-$ which makes it suitable for a perturbation expansion contained in the $\mr$ subspace. Wherever the screened Coulomb interaction enters the perturbation expansion, we use the partially screened Coulomb interaction $W_{\mr}$ to avoid double-counting correlation. The cRPA is already established as an effective tool in strongly-correlated physics and quantum embedding \cite{aryasetiawan_prb_70,sasioglu_prb_83,biermann_prb_86,biermann_prb_96,werner_prb_91,biermann_prb_96}. The physics of Eq.~\ref{excitation} is an effective Hamiltonian with a one-body part that is $GW_{\mr}$ quasiparticles and their two-body interaction via $W_{\mr}$.

With $\Omega^{\mr}$ and $E_0^{\mr}$ (described elsewhere \cite{dvorak_prb}) in hand, we can insert $H^{\mr}$ in place of $\mr H \mr$ in Eq.~\ref{eigen}. The final effective equations, demonstrated in Fig.~\ref{embedding}b, are
\begin{eqnarray}
M_{II'}(\omega) = \sum_J \bra{I} &H& \ket{J} \frac{1}{( \omega - \Delta) - \Omega_J^{\mr}} \bra{J} H \ket{I'}  \nonumber  \\
 \big[ \bra{I} H \ket{I'} &+& M_{II'}(\omega) \big] \phi_{\alpha} = E_{\alpha} \phi_{\alpha} \label{final}
\end{eqnarray}
where $\omega \equiv E - E_0$ and $\Delta$, which is on the scale of a correlation energy, is related to the calculation of $E_0^{\mr}$. The matrix elements $\bra{I} H \ket{I'}$ and $\bra{I} H \ket{J}$ are computed with the exact many-body Hamiltonian using the Slater-Condon rules \cite{slater_pr_34,condon_pr_36}. For the ground state, $\omega$ is set to zero and no self-consistent iterations are needed. For excited states, the excitation energy must be found self-consistently by iterating Eq.~\ref{final} until the excitation energy, $\Omega_{\alpha} = E_{\alpha} - E_0$, equals the evaluation energy, $\omega = \Omega_{\alpha}$.

\begin{figure}
\begin{centering}
\includegraphics[width=1.0\columnwidth]{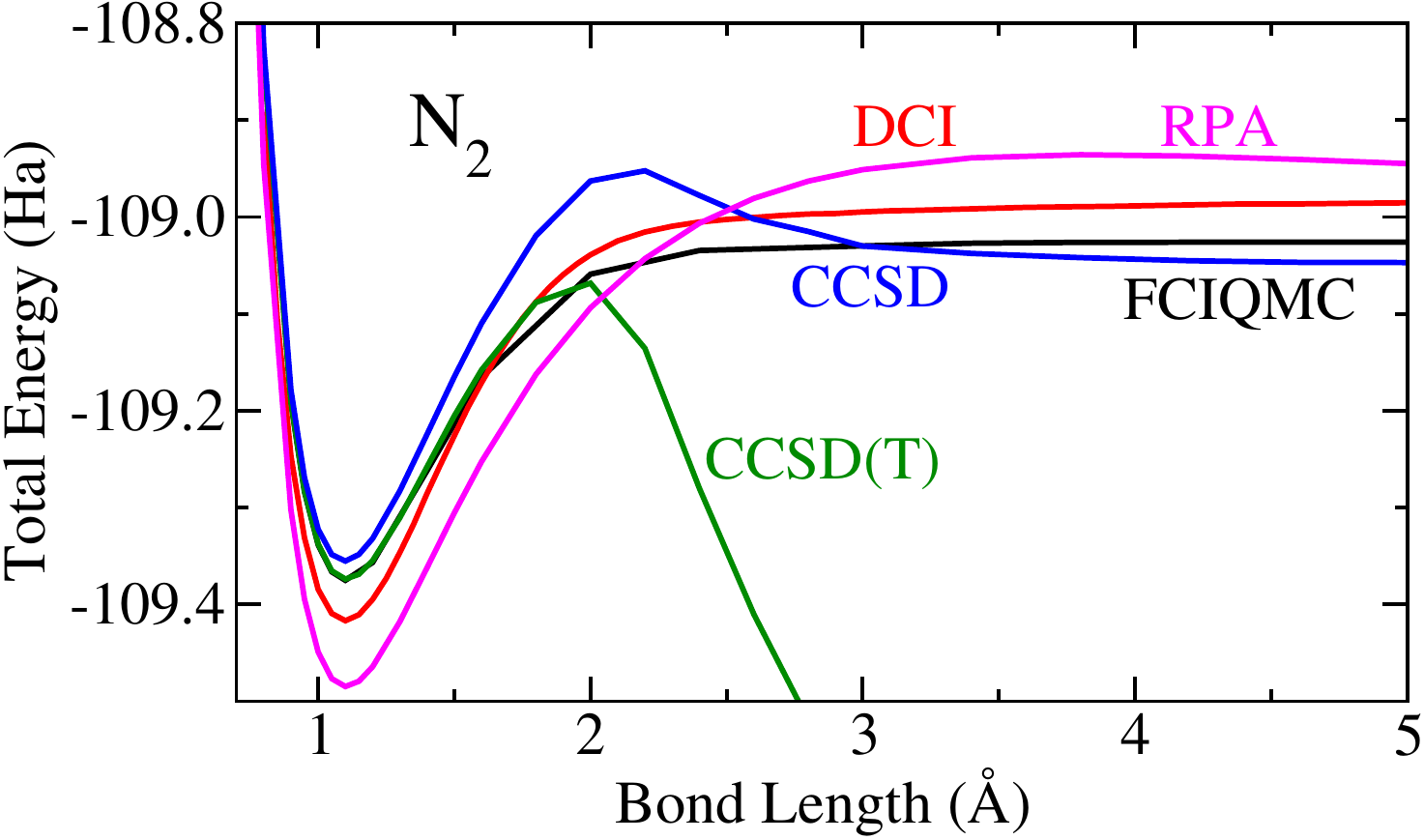}
\includegraphics[width=1.0\columnwidth]{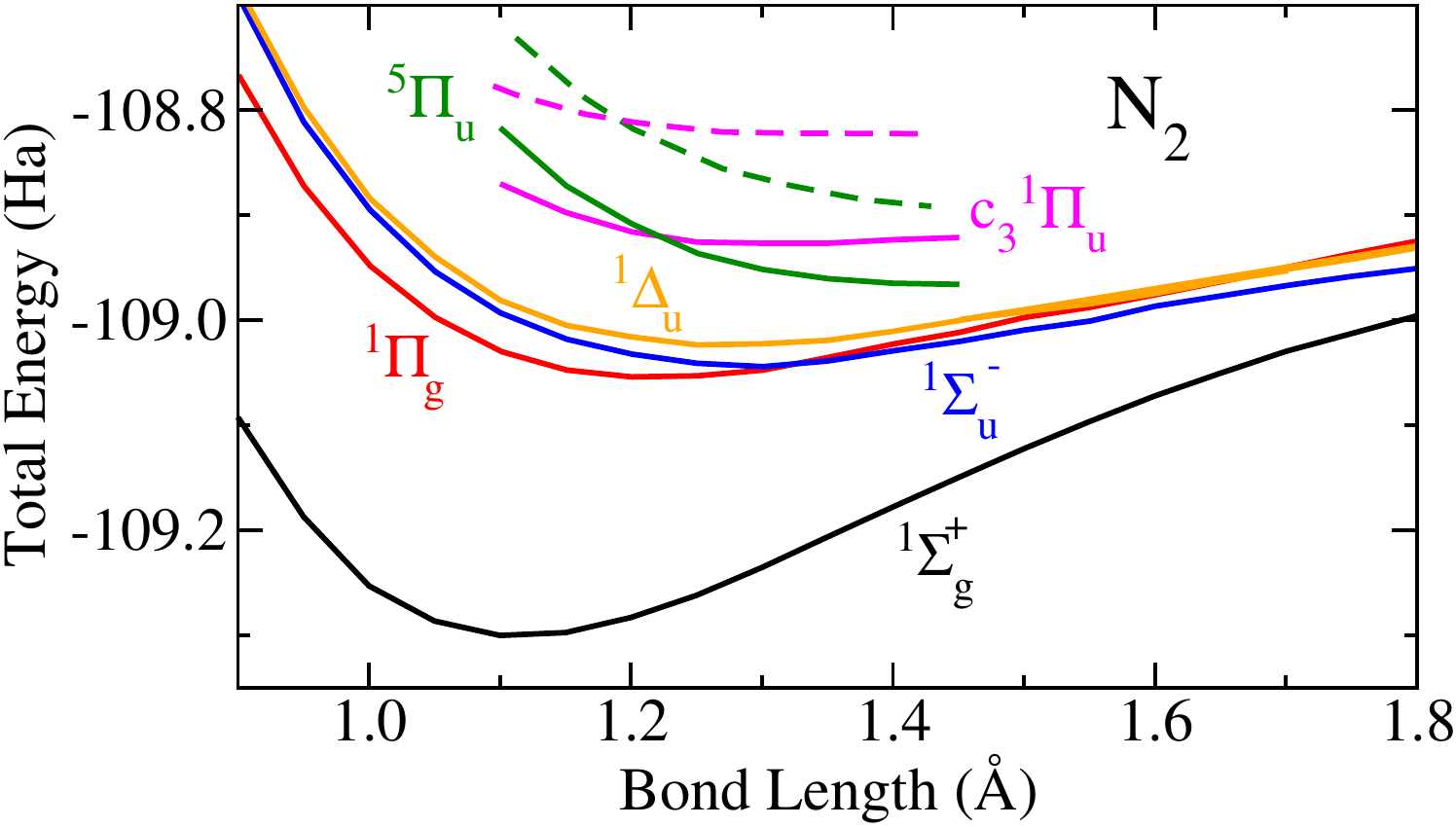}
\caption{Top: Dissociation curve of the N$_2$ dimer computed with DCI $(6,6)$ in the cc-pVTZ \cite{dunning_jcp_90} basis set compared against exact results (FCIQMC), RPA, CC with single and double excitations (CCSD), and CCSD with perturbative triple excitations (CCSD(T)). Reference data taken from Ref.~\onlinecite{zhang_prl_117}. Bottom: DCI excited state energy surfaces computed with the cc-pVDZ basis. Our DCI calculations do not use symmetry and we take the state labeling from Ref.~\onlinecite{larsen_jcp_113}. Dashed lines are FCI results from Ref.~\onlinecite{larsen_jcp_113}. \label{binding}}
\end{centering}
\end{figure}

We first test the theory by dissociating the N$_2$ dimer in the triple bond AS. Bond breaking of molecular dimers is a challenging multi-reference problem because the correct ground state wave function cannot be written as a single Slater determinant \cite{szabo_qchem,larsen_jcp_113,gagliardi_jcp_134,olsen_jcp_140}. We perform DCI calculations by exactly diagonalizing the $(6,6)$ AS (6 electrons distributed in 6 spatial orbitals) dynamically embedded in the full set of molecular orbitals. Our calculations based on FHI-AIMS \cite{blum_cpc_180,levchenko_cpc_192,ihrig_njp_17,ren_njp_14,Caruso/etal:2013_2} always use a restricted Hartree-Fock (RHF) starting point with $G_0W_{0,\mr}$@RHF in the basis of RHF orbitals. Fig.~\ref{binding} shows our DCI results compared to two versions of coupled cluster (CC), the random phase approximation (RPA), and full configuration interaction quantum Monte Carlo (FCIQMC). Our DCI calculation is free of unphysical bumps or divergences in the dissociation curve characteristic of single-reference methods. The overall agreement with high level results is satisfactory considering the relative ease of our augmented $(6,6)$ CI calculation.

Continuing with the challenging case of N$_2$, we compute excited state energy surfaces along the dissociation path, shown in Fig.~\ref{binding}. Qualitatively, the ground state and three lowest excited states closely match FCI results \cite{larsen_jcp_113}. Our primary interest is with the conical intersection between the higher energy $^5\Pi_u$ and $c_3\! ^1\Pi_u$ states near 1.3 \AA. FCI results of this intersection from Ref.~\onlinecite{larsen_jcp_113} are shown in Fig. \ref{binding} with dashed lines. There is a vertical shift between DCI and FCI data due to the overestimate of dynamic correlation, but the shape of the DCI intersection agrees with the FCI results. This intersection is missed by all variants of CC tested in Ref.~\onlinecite{larsen_jcp_113}. Properly describing the conical intersection demonstrates that DCI is unbiased towards any single \sd configuration and can treat near degeneracies among multi-configurational states.

For a quantitative comparison, we report equilibrium excitation energies in Table \ref{vertical}. We expect excitation energies to be the major strength of the theory. Systematic errors in total energies for both ground and excited states may cancel during internal $\omega$ iterations to compute $\Omega$. Our DCI calculation for excited states of N$_2$ shows good agreement with experiment and equation-of-motion CC (EOM-CCSD), and noticeably improves upon $GW$/BSE \cite{hirose_prb_91,head-gordon_jcp_103}. The unusual bonding of the carbon dimer is a challenging problem for many theories \cite{shaik_nc_4,pavlyukh_prb_75}, and the $\pi \rightarrow \pi^*$ transitions in ethene and butadiene are additional benchmark tests in quantum chemistry with high quality theoretical data for comparison \cite{alavi_jctc_8,watts_jcp_105,lischka_jcp_110,lischka_tca_112,chan_jctc_8}. The lowest excitation energies for C$_2$, ethene, and butadiene calculated with DCI, shown in Tables \ref{vertical} and \ref{organics}, are in excellent agreement with benchmark theory. For butadiene and ethene, our discrepancy with experiment 
can be attributed to an incomplete basis and nonadiabatic coupling present in experiment \cite{alavi_jctc_8}. In butadiene, for which we find the many-body excitation to have one dominant single excitation, our computed excitation energy changes by $<0.05$ eV by varying the active space from $(2,2)$ to $(8,8)$ \cite{supp_info}. Even with a small AS, DCI can describe such well-defined excitations $-$ the effects of configurations surrounding the dominant one are already captured by our quasiparticle Hamiltonian.
\begin{table}
\caption{Vertical singlet excitation energies (eV) of N$_2$ \cite{hirose_prb_91} and C$_2$\footnote{We perform our own calculation for C$_2$ at the $G_0W_0$@HF/BSE level. The N$_2$ value from Ref.~\onlinecite{hirose_prb_91} is based on $G_0W_0$@LDA/BSE.} computed with the Bethe-Salpeter equation ($GW$/BSE), EOM-CCSD, \cite{hirose_prb_91,head-gordon_jcp_103} and DCI. Our $(6,6)$ and $(8,8)$ DCI calculations are performed at the experimental bond lengths of 1.0977 \AA $\,$ and 1.2425 \AA $\,$, respectively for N$_2$ and C$_2$, in the cc-pVQZ basis. \label{vertical}}
\begin{ruledtabular}
\begin{tabular}{ c c c c c }
       & $GW$/BSE & EOM-CCSD & \textbf{DCI} & Exp. \cite{oddershede_cp_97}  \\
   \hline
  N$_2$  & 7.93  & 9.47 & \textbf{9.33} & 9.31    \\
  C$_2$  & $<0.1$ & 1.33 & \textbf{1.11}  & 1.23    \\
\end{tabular}
\end{ruledtabular}
\end{table}
\begin{table}
\caption{Vertical singlet excitation energies (eV) of ethene (C$_2$H$_4$) and butadiene (C$_4$H$_6$) computed at their experimental \cite{herzberg} and MP2 \cite{alavi_jctc_8} geometries in the cc-pVTZ and cc-pVDZ basis sets, respectively. We use DCI active spaces of (6,6) and (4,4), respectively, to correlate the $\pi \rightarrow \pi^*$ transition. For ethene, our (6,6) AS includes low energy $\sigma$ and $\sigma^*$ states. Reference CCSDT and FCIQMC data are in ANO-L-VXZP (X=D or X=T) basis sets \cite{alavi_jctc_8}. \label{organics}}
\begin{ruledtabular}
\begin{tabular}{ c c c c c }
      & EOM-CCSDT & FCIQMC & \textbf{DCI} & Exp. \cite{ethene_exp_1,ethene_exp_2,butadiene_exp_1,butadiene_exp_2,butadiene_exp_3}  \\
      \hline
  C$_2$H$_4$   & 7.97 & 7.97 & \textbf{7.99} & 7.66           \\  
  C$_4$H$_6$   & 6.50 & 6.53 & \textbf{6.48} & 5.92           \\   
\end{tabular}
\end{ruledtabular}
\end{table}

We also consider charged excitations, which depend on two separate SCF and $G_0W_{0,\mr}$ calculations, to test the robustness of the theory. The ionization spectrum of benzene is a difficult prediction in MBPT that is sensitive to self-consistency, vertex corrections, and mean-field starting points \cite{louie_prb_94,ren_prb_92}. To describe the first 5 ionization energies, we use an orbital AS of $(10,7)$ and $(9,7)$ for the neutral molecule and ion, respectively. For such a small AS in a system as large as benzene, the correlation treatment in \sr is very important and presents a difficult test of the theory.

\begin{figure}
\begin{centering}
\vspace{0.5cm}
\includegraphics[width=1.0\columnwidth]{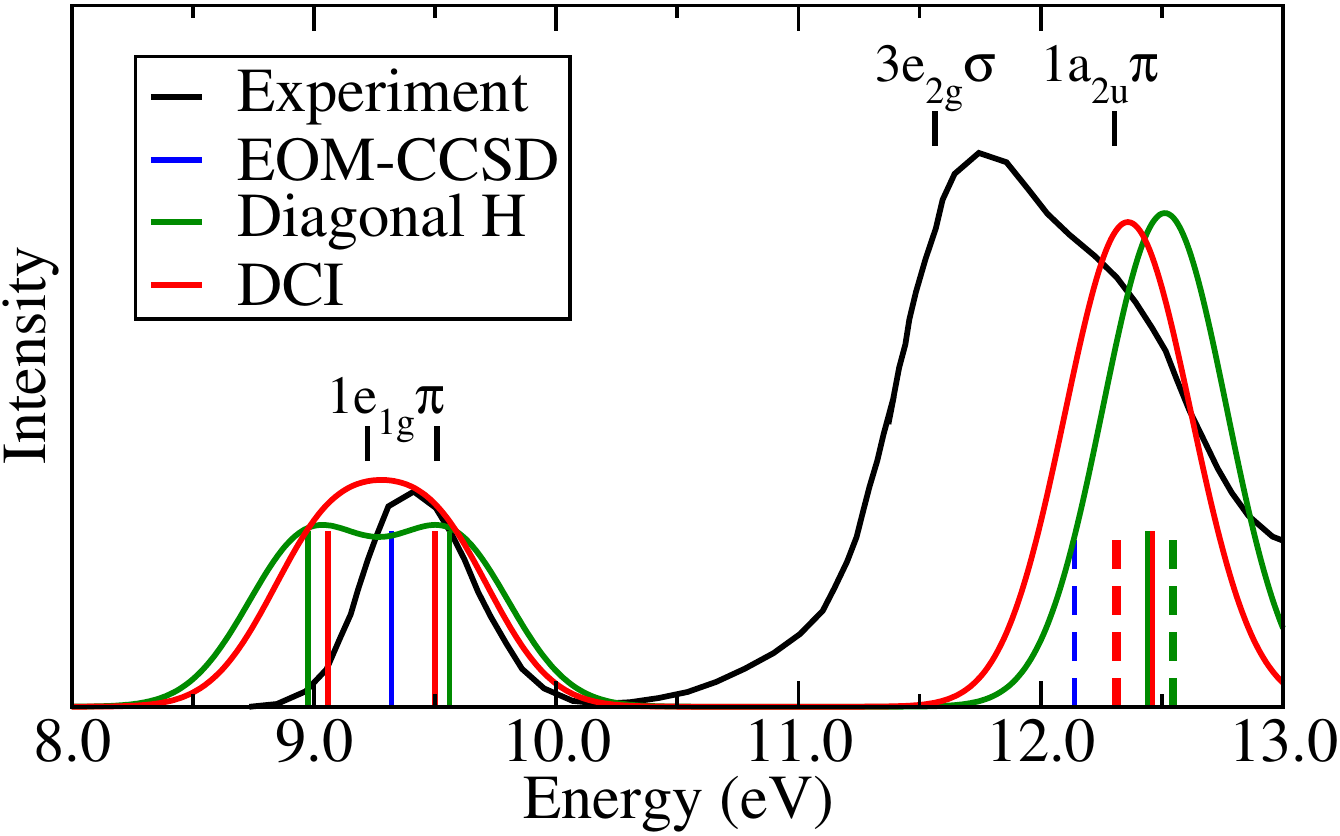}
\caption{Ionization spectrum of benzene measured by experiment \cite{liu_jpca_115} (black), computed with EOM-CCSD \cite{berkelbach_jctc_14} (blue), the Diagonal H approximation (green), and DCI (red). Peak assignments are taken from Ref.~\onlinecite{carlson_jcp_86}. We use the cc-pVDZ basis and generate \sd up to triple excitations, DCI-SDT. $\pi$ states are indicated with solid lines while $\sigma$ states are shown with dashed lines. \label{benzene} } 
\end{centering}
\end{figure}

Our DCI prediction is shown in Fig. \ref{benzene}. The first ionization potential (IP), a bonding $\pi$ state near 9 eV, is in good agreement with experiment and past results. It is encouraging that the theory can describe such a charged excitation. The $\pi$ state near 12.5 eV is also in good agreement with experiment. We predict the first $\sigma$ state to be $\sim$0.15 eV below the closest $\pi$ state. While this peak position is not perfectly aligned with experiment, our result is in good agreement with recent EOM-CCSD results \cite{berkelbach_jctc_14} ($<0.2$ eV) and the renormalized singles $GW$ approach (RS$GW$) \cite{jin_jcpl_447} without any adjustable parameters. The remaining discrepancy between theory and experiment for the $\sigma$ state of benzene could be partly due to non-adiabatic effects or, in our case, an error of the underlying $G_0W_{0,\mr}$ calculation.

For benzene, we test a number of other approximations to \sr correlation, both with and without the quasiparticle approximation. Certain approximations, such as the frozen core approximation in which \sr correlation is zero, perform so poorly that their spectra do not even resemble the experimental one. A $GW$-RPA-like approximation, in which the quasiparticle excitation energy is a sum of $GW_{\mr}$ quasiparticles without any interquasiparticle interactions, also performs very poorly (not shown here). Among these other approximations, the one with the best agreement with experiment is to use the diagonal matrix $\mr H \mr$ without any quasiparticle renormalization, which we denote ``Diagonal H" in Fig.~\ref{benzene}. In Diagonal H, the excitation energy has the same form as Eq.~\ref{excitation}, but the self-energy is the bare exchange and the screened Coulomb interaction between excited particles is instead unscreened. Additionally, the ground state energy $E_0^{\mr}$ is replaced by the energy of the reference configuration, $E^{\mathrm{ref}}$. By comparison with DCI, we see the effect of screening in the \sr subspace. For benzene, the screening effects included in DCI improve the splitting of the first $\pi$ states and the position of the $\sigma$ peak. The Diagonal H approximation reverses the ordering of the higher $\pi$ and $\sigma$ states, in worse agreement with experiment than DCI. The improved agreement with experiment by including screening with DCI gives us confidence that a quasiparticle treatment beyond $G_0W_0$@HF in the future will further improve the results.

\begin{figure}
\begin{centering}
\includegraphics[width=1.0\columnwidth]{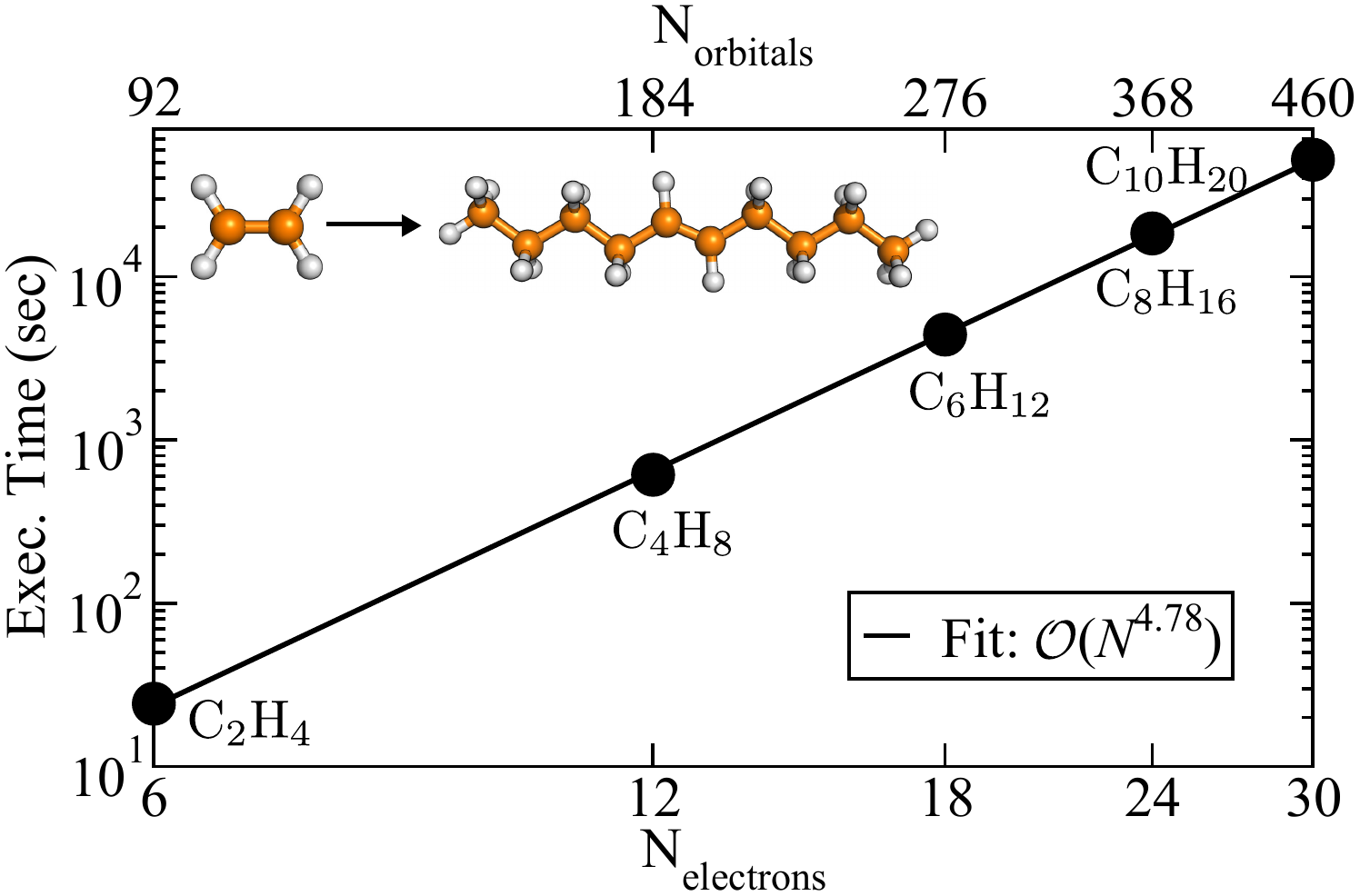}
\caption{Total time to compute the DCI Hamiltonian for alkene chains as a function of electron number, $N_{\mathrm{electrons}}$, or number of orbitals, $N_{\mathrm{orbitals}}$. \label{scaling}}
\end{centering}
\end{figure}

Finally, we discuss the computational scaling and algorithm behind our approach. Eq.~\ref{eigen} is equivalent to ED and does not, by itself, improve the computational scaling of the many-body problem. However, our diagonal quasiparticle approximation reduces scaling of the DCI Hamiltonian for a fixed AS to a much more efficient $\mathcal{O}(N^5)$. A number of important problems, including point defects in solids or $d$-electron complexes, can be formulated as a fixed AS coupled to varying bath degrees of freedom, represented by different solids or molecular ligands in these examples. To demonstrate this principle, consider a series of alkene chains of increasing length with a single double bond at their centers. For an AS correlating the double bond, DCI provides a proper multi-reference treatment of strong correlation while adding dynamic correlation with favorable $\mathcal{O}(N^5)$ scaling, as shown in Fig.~\ref{scaling}. The DCI algorithm is conceptually simple and well-suited to parallelization. The eigenvalues $\epsilon_i^{GW_{\mr}}$ and matrix elements of $W_{\mr}$ in Eq.~\ref{excitation} are precomputed numbers that never need to be updated during self-consistent iterations.

In conclusion, we have presented a new quantum embedding theory that effectively embeds a wave function calculation inside of a many-body Green's function calculation to capitalize on the strengths of both theories. Our DCI theory merges aspects of quantum chemistry, strongly-correlated physics, and $GW$ theory to provide a balanced, multi-disciplinary description of electronic correlation. Initial calculations for dimers, linear organics, and benzene demonstrate the versatility of the theory for describing different regimes of correlation.

\FloatBarrier
\vspace{0.5cm}

This work is supported by the Academy of Finland through grant Nos. 284621, 305632, 316347, and 316168. The authors acknowledge the CSC-IT Center for Science, Finland, for generous computational resources and the Aalto University School of Science ``Science-IT'' project for computational resources. The authors acknowledge A. Harju for early discussions on the topic, as well as fruitful discussions with S. Biermann, R. van Leeuwen, and Y. Pavlyukh.

\end{document}